\documentclass[aps,prd,superscriptaddress,onecolumn,showpacs,showkeys]{revtex4}
\usepackage{amssymb,amsmath,epsfig}

\usepackage{eurosym}
\usepackage{amsfonts}
\usepackage{array}
\usepackage{amsthm}
\usepackage{bm}
\usepackage{palatino}
\usepackage{mathpazo}
\usepackage{changes}
\usepackage{amssymb}
\usepackage{eurosym}
\usepackage{amsmath}
\usepackage{epsfig}
\usepackage{graphics}
\usepackage{color}
\usepackage{graphicx}
\begin{document}

\title{Effect of the quintessential dark energy on weak deflection angle by Kerr-Newmann Black hole}

\author{W. Javed} \email{wajiha.javed@ue.edu.pk}
\affiliation{Department of Mathematics, University of Education,\\
Township, Lahore-54590, Pakistan.}

\author{J. Abbas}
\email{jameelaabbas30@gmail.com}
\affiliation{Department of Mathematics, University of Education,\\
Township, Lahore-54590, Pakistan.}

\author{A. \"{O}vg\"{u}n}
\email{ali.ovgun@emu.edu.tr}
\homepage[]{https://www.aovgun.com}
\affiliation{Physics Department, Faculty of Arts and Sciences, Eastern Mediterranean
University, Famagusta, 99628 North Cyprus, via Mersin 10, Turkey}

\begin{abstract}
In this work, we study the weak gravitational
lensing in the background of Kerr-Newman black
hole with quintessential dark energy. Initially,
we compute the deflection angle of light by charged black hole with quintessential dark energy
by utilizing the Gauss-Bonnet theorem. Firstly, we
suppose the light rays on the equatorial plane in
the axisymmetric spacetime. In doing so, we first
find the corresponding optical metrics and then
calculate the Gaussian optical curvature to utilize
in Gauss-Bonnet theorem. Consequently, we calculate
the deflection angle of light for rotating
charged black hole with quintessence.
Additionally, we also find the
deflection angle of light for Kerr-Newman black
hole with quintessential dark energy. In order
to verify our results, we derive deflection
angle by using null geodesic equations which reduces to the deflection angle of Kerr
solution with the reduction of specific
parameters. Furthermore, we analyze the
graphical behavior of deflection angle
$\Theta$ w.r.t to impact parameter $b$. Our
graphical analysis retrieve various
results regarding to the deflection angle by the Kerr-Newman
black hole with quintessential dark energy.
\end{abstract}

\date{\today}

\maketitle

\section{Introduction}

Einstein general theory of relativity is based on numerous experimental and
observational tests \cite{R31}. These tests revealed that the experimental data
is compatible with theoretical predictions of this theory, i.e., some of the 
astrophysical observations \cite{R32} like gravitational
waves, gravitational lensing, black holes (BHs), wormholes etc. Einstein assumed
a universe shape as a single fabric sheet, hewn from space and time. According
to this theory, the force of gravity is the result of curvature in space-time and
gravitational waves are ripples in the fabric of a universe that are produced by
the collision of the massive objects such as BHs. Gravity bends light such phenomenon
is known as \textit{gravitational lensing}. The gravitational lensing is as the distribution of
matter between a source and observer, which is capable of bending of light from
the source to the observer and the bending of light is one of the prediction of Einstein's theory.
Gravitational lensing particularly plays a vital role in the study of distribution of dark
matter in universe and for imaging the most distant galaxies. The prediction that light bends
when passing near massive objects is valid in Newtonian physics but is more significant in general
theory of relativity. The gravitational lensing is very helpful for the detection of the BHs in the
universe and this phenomenon indicates that light rays get deflected due to the curvature of spacetime.
The gravitational lensing has been classified in the literature as a strong lensing and weak lensing.
Strong lensing helps us to find the magnification, position and time delays of the images by BHs.
Strong lensing phenomena required in many cases which yields more information from experimental frame of
reference to analyze other exotic objects such that monopoles \cite{R24}, boson stars \cite{R25,R26}, fermion stars
\cite{R27,R28} etc. While the weak gravitational lensing provide a way to find the mass of astronomical
objects without requiring about their composition or dynamical states. Weak lensing also investigates the cause
of the accelerated expansion of the universe and also distinguish between modified gravity and dark energy (DE).
Recently, scientists detected the event horizon of
BH by the reference of Event Horizon Telescope (EHT) \cite{Akiyama:2019cqa}; that's why this concept has claimed a
great attention and many astronomer concentrate on it to acquire exact results \cite{R29,Konoplya:2019sns,Bambi:2019tjh,Shaikh:2019fpu,Abdikamalov:2019ztb,Abdujabbarov:2017pfw,Abdujabbarov:2016efm,Abdujabbarov:2016hnw,Abdujabbarov:2015pqp,Papnoi:2014aaa,Atamurotov:2013sca}.

Over the past many years ago, numerous complimentary experimental
investigations on cosmological measures have detected powerful indications to speed up extension over the cosmos: distance measurements
of distant model Ia supernovae \cite{j15}, the observations of the cosmic
microwave framework of anisotropies additionally the neutron phonic
oscillations in the matter spectrum eliminated
into the universe lists. In taking up general theory of relativity, a negative pressure
segment was suggested as a most viable explanation to the acceleration found. The best approach to render like a method, though it endures with unresolved issues for instance the finishing touches and
cosmic concurrence issues is the cosmological constant $\Lambda$, among a constant state-equation (EoS) parameter $\omega=-1$. Dynamic DE patterns, suchlike the
quintessence, phantom, quintom and Ricci dark energy (RDE) patterns,
have also been intended into the documentation \cite{j16},\cite{j17}.

The presence of a high range of hypothetical patterns which does not 
conflict by means of the fundamental discovery connected with rapid extension has 
motivated a series of observational analysis, centered on, for 
example, angular size-redshift data from condensed radio sources, 
age-redshift interaction, galaxy cluster lookback time, galaxy 
cluster x-ray luminosity, and Hubble parameter results. In this sense, strong gravitational lensing plays a vital role , offering cosmological evaluations for instance gravitational lensing statistics, Einstein rings inside the celestial body cosmos networks, galaxy blocks which act by way of lens on major red shift regimes in the context, and time delay estimates.

The intriguing form of energy with positive energy density 
and large negative pressure is represented as \textit{dark energy}
\cite{R1}-\cite{R3}. There are two different approaches to deal with DE 
such as cosmological constant and quintessence parameter. Cosmological 
constant (denoted as Greek word $\Lambda$) describes the energy density of
space \cite{R4,R1}. According to some physicists, it
has been proposed as a fifth fundamental force. Quintessence differs from
cosmological constant as it changes over time due to that it is dynamic,
otherwise the cosmological constant by definition does not change \cite{R4.1}.
Moreover, $\omega$ is an equation of state parameter which divides DE phase in
different eras such that DE phase consists of quintessence era for 
$-1<\omega\leq\frac{-1}{3}$ whereas $\omega=-1$ analogous to cosmological constant.

In doing so, Gibbon and Werner \cite{R8} presented a method to derive deflection 
angle of light rays by using Gauss Bonnet theorem (GBT).
This method is use to find the integral in a finite domain bounded by the light ray.
Then, Werner has enlarged this formalism and derive
the angle of deflection by a Kerr-Newman BH by using the Nazim's method
with Rander-Finsler metric \cite{R5}. Recently, Ishihara et al. \cite{R6,Arakida:2017hrm,Ono:2018ybw,Ono:2017pie}
computed the deflection angle in a static, spherically symmetric and asymptotically
flat space by using the finite distance from an observer to a light source. Moreover,
the GBT has been stretched out to the wormhole geometries and non-asymptotically flat
spacetime with topological effects \cite{Jusufi:2017vta,Ovgun:2018xys,Jusufi:2017mav}. A very important contribution has been
currently made by Jusufi and \"{O}vg\"{u}n \cite{Jusufi:2017vew} who discussed about the quantum correction
effects on the deflection of light by quantum improved Kerr BH pierced in cosmic string.
Stunningly, this method was shown in a suitable way to calculate the deflection angle in
spacetimes with topological defects by cosmic strings
and global monopoles \cite{Jusufi:2017lsl,Jusufi:2017hed,Ono:2018jrv,Ovgun:2019wej}. This method has been utilized in various papers for 
different types of spacetimes \cite{Crisnejo:2018uyn,Sakalli:2017ewb,Ovgun:2018prw,Jusufi:2017uhh,Jusufi:2018jof,Ovgun:2018fnk,Ovgun:2018ran,Ovgun:2018oxk,Ovgun:2018fte,Ovgun:2018tua,Javed:2019qyg,Javed:2019a,Javed:2019b}.

Xu and Wang \cite{22} examined the Kerr-Newman BH in the background of
quintessential field \cite{Lammerzahl:2018zvb,Visser:2007fj}. It is considerable that the Newman-Janis algorithm
does not involve cosmological constant, therefore they enlarge the Kerr-
Newman solution to the Kerr-Newman-AdS solution with quintessential dark
energy through direct computations which satisfy the Einstein Maxwell field
equations in quintessential dark energy with negative cosmological constant.
Moreover, they also investigated the singularity of Kerr-Newman-AdS BH by
quintessence dark energy.

In this continuation, we are going to extend this formalism for charged BH in quintessential dark energy and also for Kerr-Newman
BH with quintessence. In our analysis, we focus on a non-singular
domain outside of the light ray. As for asymptotically flat space,
deflection angle $\Theta$ can be computed as follows \cite{R8}:
\begin{equation}
\Theta=-\int\int_{D_{\infty}}\mathcal{K}d\tilde{\sigma},\nonumber
\end{equation}
where $\mathcal{K}$ represents the Gaussian optical curvature and $d\tilde{\sigma}$ stands for
surface element of optical geometry. It is remarkable that the above expression
of deflection angle just fulfill the asymptotically flat
space, while for non-asymptotically flat metric only a finite distance corrections
can be considered. We analyze the graphical behavior of deflection angle with
respect to BH charge as well as impact, quintessence and rotation parameters.
This paper is organized as follows: In section \textbf{$\textrm{2}$},
we discuss the deflection angle of non rotating charged BH with quintessence.
In section \textbf{$\textrm{3}$}, we review some
basic concepts about Rander-Finsler geometry and compute the Gaussian optical curvature for
deflection angle of Kerr-Newman BH with quintessence. 
In section \textbf{$\textrm{4}$}, we calculate the deflection angle for Kerr-Newman BH with quintessence
by using GBT. In section \textbf{$\textrm{5}$}, we discuss the geodesic equations with the background geometry of Kerr-Newman BH
involving quintessential dark energy and compute deflection angle. Section \textbf{$\textrm{6}$} is
based on graphical analysis, we discuss deflection angle w.r.t various physical parameters.
The last section comprises of concluding remarks and results obtained from graphical analysis.

\section{ Charged Black hole with Quintessential Dark Energy}
There exist a literature related to charged BH in the framework of quintessential dark energy \cite{r21} also the metric functions are explicitly described in \cite{r22}.
The spacetime metric of charged black hole with quintessential dark energy is \cite{r22}
\begin{equation}
    d s^{2}=-F(r) d t^{2}+F(r)^{-1} d r^{2}+r^{2} d \theta^{2}+r^{2} \sin ^{2} \theta d \phi^{2} \end{equation}
    
    and then we set
the metric into equilateral plane with $\theta=\frac{\pi}{2}$, thus the corresponding expressions are
    \begin{equation}
    d s^{2}=-F(r)dt^{2}+F(r)^{-1} d r^{2}+r^{2} d \phi^{2}, \end{equation}
    where
    \begin{equation}F(r)=1-\frac{2M}{r}+\frac{Q^{2}}{r^2}-\frac{\alpha}{r^{1+3\omega}},\end{equation}
   where the $\alpha$ is the quintessence parameter
that represents the intensity of the quintessence field related to the black hole, $M$ is the black hole mass, and $\omega$ is the parameter describing the equation of state with $\omega= p/\rho$, where $p$ and $\rho$ are the pressure and energy
density of the quintessence respectively. Note that $\omega$ can not be equal to $0$, $1/3$, $−1$. The parameter $\omega$ is $-1 < \omega < −1/3$, then it can explain the accelerating expansion of universe. The relation between parameters are given as: $\omega=\frac{1}{3}\hat{\alpha}$ \cite{r22} . The optical metric is
    \begin{equation}
    d t^{2}=\frac{1}{F(r)^{2}} d r^{2}+\frac{r^{2}}{F(r)} d\phi^{2}. \end{equation}

Then we calculate the Gaussian curvature of the optical charged black hole as follows:
\begin{equation}
   \mathcal{K}=\frac{R_{icciScalar}}{2},
\end{equation}
after simplification the Gaussian optical curvature in weak field limit is stated as 
\begin{eqnarray}
\mathcal{K}&\approx& -\frac{2M}{r^{3}}+\frac{3Q^{2}}{r^{4}}-\frac{6MQ^{2}}{r^{5}}
+\frac{1}{2r^{6}}\left[-(9\omega^{2}+9\omega+2)r^{3-3\omega}\right.\nonumber\\&+&\left.
18M(\omega^{2}+\frac{2}{3}\omega+\frac{1}{3})r^{2-3\omega}-9Q^{2}(\omega^{2}+
\frac{1}{3}\omega+\frac{2}{3})r^{1-3\omega}\right]\alpha.
\end{eqnarray}
Now, by using the GBT and Eq. $(6)$, the deflection angle of charged BH with quintessential dark energy is obtained as
\begin{equation}
\Theta\approx \frac{4M}{b}+\frac{3\pi Q^{2}}{4b^{2}}+\frac{2\alpha}{b}-\frac{3\alpha\omega}{b}+
\frac{6\alpha\omega}{b}(\ln(2)-\ln(b))+\mathcal{O}(\alpha^{2},M^{2},Q^{3}).
\end{equation}
In the above expression the term $\mathcal{O}(\alpha^{2},M^{2},Q^{3})$ described that we only consider the linear order term of mass function, quintessence parameter and second order for BH charge also ignore the higher order terms.
It is to be noted that the mass term decreases the deflection angle while quintessence parameter $\alpha$ increases the deflection angle. On the other hand, the deflection angle has an indirect relation with impact parameter $b$.

\section{Graphical Analysis}
This section is devoted to discuss the graphically behavior
of deflection angle $\Theta$. We also demonstrate the physical
significance of these graphs to analyze the impact of BH charge $Q$
and quintessence $\alpha$ on deflection angle by examining the stability
and instability of BH.
\subsection{Deflection angle with Impact parameter $b$}
This subsection is based on the analysis of deflection angle $\Theta$
with impact parameter $b$ for different values of BH charge $Q$,
and quintessence parameter $\alpha$ for fixed $M=1$ and $\omega=-\frac{2}{3}$.
\begin{center}
\epsfig{file=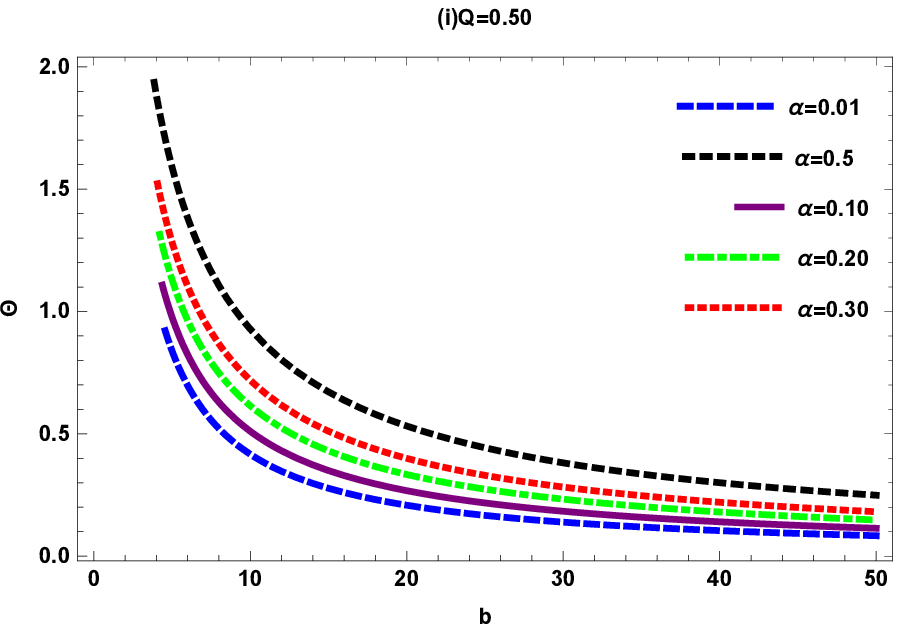,width=0.50\linewidth}\epsfig{file=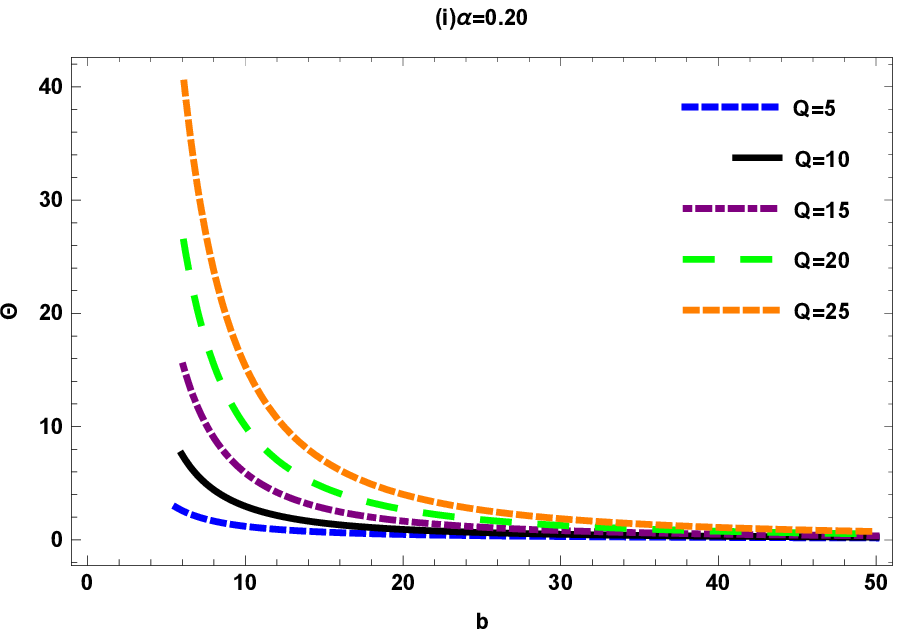,width=0.50\linewidth}\\
{Figure 1: Relation between $\Theta$ and $b$}.
\end{center}
\begin{itemize}
\item \textbf{Figure 1} shows the behavior of $\Theta$ w.r.t impact parameter $b$.
\begin{enumerate}
\item In figure (i), This plot shows the behavior of $\Theta$ with $b$ for fixed $Q$
 and varying $\alpha$. We analyzed that deflection angle initially exponentially
 decreasing and then goes to positive infinity. We also observed that for $0.01\leq\alpha \leq1$,
  we obtain the stable behavior. For negative values of $\alpha$, the deflection angle is negatively
  decreasing which shows the unstable state of BH. Hence, we analyzed the stable state of BH only
  for positive $\alpha$.

 \item In figure (ii), This plot represents the behavior of $\Theta$ with $b$ for fixed $\alpha$
  and varying $Q$ we observed that the deflection angle is gradually decreasing and
  then goes to positive infinity.
\end{enumerate}
\end{itemize}

\section{Rander-Finsler Geometry: Kerr-Newman Black Hole}

In this section, we illustrate the charged rotating Kerr-Newman BH in the background of quintessential dark energy. There exist a literature related to this BH recently discussed by Javed and Babar \cite{R101}.
For this purpose, we consider the line-element of general Kerr-Newman-AdS BH in the back ground of quintessential
dark energy, i.e., \cite{Xu:2016jod,j12}
\begin{eqnarray}
ds^{2}&=&\frac{\Sigma^{2}}{\Delta_{r}}dr^{2}+\frac{\Sigma^{2}}{\Delta_{\theta}}d\theta^{2}
+\frac{\Delta_{\theta}\sin^{2}\theta}{\Sigma^{2}}\left(a\frac{dt}{\Xi}-(r^{2}+a^{2})
\frac{d\phi}{\Xi}\right)^{2}\nonumber\\&-&\frac{\Delta_{r}}{\Sigma^{2}}
\left({dt}{\Xi}-a\sin^{2}\frac{d\phi}{\Xi}\right)^{2},\label{J1}
\end{eqnarray}
where
\begin{eqnarray*}
\Delta_{r}&=&r^{2}-2Mr+a^{2}+Q^{2}-\frac{\Lambda}{3}r^{2}(r^{2}+a^{2})-\alpha r^{1-3\omega},\nonumber \\
\Delta_{\theta}&=&1+\frac{\Lambda}{3}a^{2}\cos^{2}\theta,~~~~
\Xi=1+\frac{\Lambda}{3}a^{2},~~~~
\Sigma^{2}=r^{2}+a^{2}\cos^{2}\theta,
\end{eqnarray*}
where $M$ is the mass, $Q$ is BH charge, $a$ represents spin and $\alpha$ is
quintessence parameter representing dark energy. For $\Lambda=0$, this solution reduces to a charged
rotating BH solution in the presence of quintessential dark energy.
Thus, for the sake of simplicity, we consider $\Lambda=0$ and also set
the metric into equilateral plane with $\theta=\frac{\pi}{2}$, thus the corresponding expressions are 
\begin{eqnarray}
\Delta_{r}&=&r^{2}-2Mr+a^{2}+Q^{2}-\alpha r^{1-3\omega},\nonumber \\
\Delta_{\theta}&=&1,~~~ \Xi=1,~~~ \Sigma^{2}=r^{2}.
\end{eqnarray}
It is appropriate to first find the black hole optical metric by imposing the null condition $ds^{2}=0$,
and solving the space-time metric for $dt$, which yield the generic form as
\begin{equation}
dt=\pm\sqrt{\xi_{ij}(x)v^{i}v^{j}}+\gamma_{i}(x)v^{i},
\end{equation}
moreover, the expression $\xi_{ij}(x)v^{i}v^{j}$ can be derived as follows
\begin{widetext}
\begin{equation}
\xi_{ij}(x)v^{i}v^{j}=\frac{r^{4}}{\Delta_{r}(\Delta_{r}-a^{2})}dr^{2}+\left(\frac{(r^{2}+a^{2})-a^{2}\Delta_{r}}
{\Delta_{r}-a^{2}}+\frac{a^{2}((r^{2}+a^{2})+\Delta_{r})^{2}}{\Delta_{r}-a^{2}}\right)d\phi^{2},\nonumber
\end{equation}
\end{widetext}
while, $\gamma_{i}(x)v^{i}$ can be expressed as
\begin{equation}
\gamma_{i}(x)v^{i}=-\frac{a(r^{2}+a^{2}-\Delta_{r})}{\Delta_{r}-a^{2}}d\phi.\nonumber
\end{equation}
The deformed form of the given metric in Finsler-Randers type can be expressed as \cite{R5}
\begin{widetext}
\begin{eqnarray}
\tilde{\mathcal{F}}\left(r,\phi,\frac{dr}{dt},\frac{d\phi}{dt}\right)&=&\sqrt{\frac{r^{4}}{\Delta_{r}
(\Delta_{r}-a^{2})}\left(\frac{dr}{dt}\right)^{2}+\frac{(r^{2}+a^{2})-a^{2}\Delta_{r}}{\Delta_{r}-a^{2}}}\nonumber\\
&\times&\sqrt{\frac{a^{2}((r^{2}+a^{2})+\Delta_{r})^{2}}{\Delta_{r}-a^{2}}\left(\frac{d\phi}{dt}\right)^{2}}-
\frac{a(r^{2}+a^{2})}{\Delta_{r}-a^{2}}\nonumber\\&+&\frac{a \Delta_{r}}{\Delta_{r}-a^{2}}\left(\frac{d\phi}{dt}\right).
\end{eqnarray}
\end{widetext}
The Randers-Finsler optical metric $\tilde{\mathcal{F}}$ is actually use to calculate the null geodesics in our physical metric. Since $dt=\tilde{\mathcal{F}}(u,du)$,
on the other hand by Fermat's principle light rays $(\tilde{\gamma})$ are selected by the following condition
\begin{equation}
\delta\int_{\tilde{\gamma}}dt =\delta\int_{\tilde{\gamma}\tilde{\mathcal{F}}}\tilde{\mathcal{F}}(u,\dot{u})dt=0.
\end{equation}
Notice that these light rays $(\tilde{\gamma})$ are also geodesic $(\tilde{\gamma}_{\mathcal{F}})$
of the Rander metric. Thus, it is remarkable that Rander-Finsler metric $\tilde{\mathcal{F}}$
generalizes the Fermat's principle. The Rander-Finsler metric in the sense that the Hessian
\begin{equation}
g_{ij}(u,v)=\frac{1}{2} \frac{\partial^{2}\tilde{\mathcal{F}}^{2}(u,v)}{\partial v^{i}\partial v^{j}} \label{2},
\end{equation}
is positive definite, where $u\in \mathcal{M}$, $v\in T_{u}\mathcal{M}$. In order to continue along the construction of a
Riemannian manifold $(\mathcal{M},\bar{g})$ which osculate with the Randers
manifold $(\mathcal{M},\tilde{\mathcal{F}})$ by applying the so-called Nazim's method.
We consider a non-zero smooth
vector field $\bar{v}$  over $\mathcal{M}$ containing the tangent vector
with the geodesic $\tilde{\gamma}_{\tilde{\mathcal{F}}}$, stated as $\bar{v}
(\tilde{\gamma}_{\tilde{\mathcal{F}}})=\dot{u}$, i.e.,
\begin{equation}
\bar{g_{ij}}(u)=g_{ij}(u,\tilde{v}(u)).\label{1}
\end{equation}
The geometrical quantities which osculate the Riemannian manifold, for illustration
in $T_{u}\mathcal{M}$ angle between vectors can be measured by the
metric $\bar{g}$. Thus, the geodesic $\tilde{\gamma}_{\tilde{\mathcal{F}}}$
of Randers manifold is also a geodesic $\tilde{\gamma}_{\bar{g}}$ of Riemanian manifold, i.e.,
$(\tilde{\gamma}\tilde{\mathcal{F}}=\tilde{\gamma}_{\bar{g}})$ by supposition
\begin{equation}
\ddot{u}^{i}+\Gamma^{i}_{jk}(u,\dot{u})\dot{u}^{j}\dot{u}^{k}=\ddot{u}^{i}+\tilde{\Gamma}^{i}_{jk}(u)\dot{u}^{j}\dot{u}^{k}=0 ,
\end{equation}
to start with the Randers-Optical geometry $(\mathcal{M},\tilde{\mathcal{F}})$,
suppose in the equatorial plane, a region $D_{R}\subset \mathcal{M}$ which is
bounded by the light ray $\tilde{\gamma}_{\tilde{\mathcal{F}}}$ with co-ordinate
radius $b$ and a circular curve $\tilde{\gamma}_{R}$ with co-ordinate radius $R$.
These curves are as follows \cite{R5}

\begin{equation}
\tilde{\gamma}_{\tilde{\mathcal{F}}}:u^{i}(x)=\mu^{i}(x), x\in [0,n]
\end{equation}
\begin{equation}
\tilde{\gamma}_{\tilde{\mathcal{R}}} : u^{i}(x)=\xi^{i}(x), x\in [0,n^{\star}].
\end{equation}
Now, consider $\varsigma =x/n \in$ $(0,1)$ across the curve $\tilde
{\gamma}_{\tilde{\mathcal{F}}}$ and $\varsigma^{\star}=
1-x/n^{\star}\in (0,1)$ across the $\tilde{\gamma}_{\tilde{\mathcal{R}}}$ ,
each point on the geodesic $\tilde{\gamma}_{\tilde{F}}$ directly paired with
one point $\xi^{i}(n^{\star})$ on curve $\tilde{\gamma}_{\tilde{\mathcal{R}}}$
by letting $\varsigma =\varsigma^{\star}$. On the other hand, we also analyze that there
is a family of smooth curves $u^{i}(\tilde{\sigma},\varsigma)$, in such a
way that for each point pair there is precisely one curve which touches the
boundary of $\tilde{\gamma}_{\tilde{\mathcal{F}}}$ and $\tilde{\gamma}_{\tilde{R}}$.
In other words we can say that when $\mu^{i}(\varsigma)=u^{i}(0,\varsigma)$ then
$u^{i}(\tilde{\sigma},\varsigma)$ touches the boundary of $\tilde{\gamma}_{\tilde{\mathcal{F}}}$,
$\tilde{\sigma}$ represent as a new parameter. Then following equation holds
\begin{equation}
\dot{\mu}^{i}(\varsigma)=\frac{d\mu^{i}}{dx}(\varsigma)=\frac{du^{i}}{d\tilde{\sigma}}(0,\varsigma).
\end{equation}
Similarly, we can say that when $\xi^{i}(\tilde{\sigma},\varsigma)=
u^{i}(1,\varsigma)$ then $u^{i}(\tilde{\sigma},\varsigma)$ touches
the curve $\tilde{\gamma}_{\tilde{\mathcal{R}}}$ which is as follows
\begin{equation}
\dot{\xi^{i}}(\varsigma)=\frac{d\xi^{i}}{dx}(\varsigma)=\frac{du^{i}}{d\tilde{\sigma}}(1,\varsigma).
\end{equation}
In general, we can say that a smooth and non zero tangent vector field
\begin{equation}
\tilde{v}^{i}(u(\tilde{\sigma},\varsigma))=\frac{du^{i}}{d\tilde{\sigma}}(\tilde{\sigma},\varsigma),
\end{equation}
with a group of smooth curves which fulfill the accompanying relation \cite{R5} 
\begin{eqnarray}
u^{i}(\tilde{\sigma},\varsigma)=\mu^{i}(\varsigma)+\dot{\varsigma}^{i}(\varsigma)\tilde{\sigma}+\mathcal{Y}
(\varsigma)\tilde{\sigma}^{2}+\mathcal{Z}(\varsigma)\tilde{\sigma}^{3}\notag \\+\mathcal{A}^{i}(\tilde{\sigma},\varsigma)
(1-\tilde{\sigma})^{2}\tilde{\sigma}^{2},
\end{eqnarray}
where
\begin{equation}
\mathcal{Y}(\varsigma)=3\xi^{i}(\varsigma)-3\mu^{i}(\varsigma)-\dot{\xi}^{i}(\varsigma)-2\dot{\mu}^{i}(\varsigma)\mu^{i}(\varsigma),
\end{equation}
\begin{equation}
\mathcal{Z}(\varsigma)=2\mu^{i}(\varsigma)-2\xi^{i}(\varsigma)+\dot{\xi}^{i}(\varsigma)+\dot{\mu}^{i}(\varsigma).
\end{equation}
As we know that the metric we use is asymptotically plane, so for the deflection angle of the planer light ray we choose
\begin{equation}
r(\phi)=\frac{b}{\sin\phi} ,
\end{equation}
where $b$ gives the minimal radial distance of the light ray and is known as impact parameter.
Now we make the following leading terms for the vector field
\begin{equation}
\bar{v}^{r}=\frac{dr}{dx}=-\cos\phi ,\    \bar{v}^{\phi}=\frac{d\phi}{dx}=\frac{\sin^{2}\phi}{b}. \label{3}
\end{equation}
It is well-known that the vector field which we use is determined by the light
ray equation $r_{\tilde{\gamma}}$. Where our particular light ray equation shows a
straight line approximation, for this straight line approximation we verify our
angle of deflection.
\subsection{Gaussian Curvature}
In this section we compute the metric components. For this, we use Eqs. $(13)$ and $(25)$,
we get the following non-zero components
\begin{widetext}
\begin{eqnarray}
\bar{g}_{rr}&=&{\frac {{r}^{4}}{ \left( \alpha\,{r}^{1-3\,\omega}-{Q}^{2}+2\,mr-{r}^{
2} \right) ^{2}}}
,\\
\bar{g}_{r\phi}&=&2\,{\frac { \left(  \left(  \left( \cos \left( \phi \right)  \right) ^
{4}{r}^{2}+ \left( {b}^{2}-2\,{r}^{2} \right)  \left( \cos \left( \phi
 \right)  \right) ^{2}+{r}^{2} \right) r \left( {r}^{\omega} \right) ^
{3}+\bar{\Phi}\alpha\right) ma{b}^{3}\left( \cos \left( \phi \right)\right)^{3}}{ \left(  \left( \cos
 \left( \phi \right)  \right) ^{4}{r}^{2}+ \left( {b}^{2}-2\,{r}^{2}
 \right)  \left( \cos \left( \phi \right)  \right) ^{2}+{r}^{2}
 \right) ^{5/2}{r}^{2} \left( {r}^{\omega} \right) ^{3}}},\\
\bar{g}_{\phi\phi}&=&-{\frac { \left(  \left( \alpha\,{r}^{1-3\,\omega}-{Q}^{2}-{r}^{2}
 \right)  \left( \sin \left( \phi \right)  \right) ^{4}- \left( \cos
 \left( \phi \right)  \right) ^{2}{b}^{2}m \right) {r}^{4}}{ \left( 
 \left( \alpha\,{r}^{1-3\,\omega}-{Q}^{2}-{r}^{2} \right)  \left( \sin
 \left( \phi \right)  \right) ^{4}- \left( \cos \left( \phi \right) 
 \right) ^{2}{b}^{2} \right)  \left( \alpha\,{r}^{1-3\,\omega}-{Q}^{2}
-{r}^{2} \right) }},
\end{eqnarray}
with $\bar{\Phi}=2\, \left( 7/4\, \left( \cos \left( \phi \right)  \right) ^{4}{r}^
{2}+ \left( {b}^{2}-7/2\,{r}^{2} \right)  \left( \cos \left( \phi
 \right)  \right) ^{2}+7/4\,{r}^{2} \right) $.
\end{widetext}
By neglecting the higher order terms of the angular momentum $a$ we get the determinant of this metric as follows
\begin{widetext}
\begin{eqnarray}
det\bar{g}&=&-{\frac {{r}^{8} \left( {r}^{1-3\,\omega} \left( \sin \left( \phi
 \right)  \right) ^{4}\alpha- \left( \sin \left( \phi \right) 
 \right) ^{4}{r}^{2}- \left( \cos \left( \phi \right)  \right) ^{2}{b}
^{2}m \right) }{ \left( \alpha\,{r}^{1-3\,\omega}+2\,mr-{r}^{2}
 \right) ^{2} \left( {r}^{1-3\,\omega} \left( \sin \left( \phi
 \right)  \right) ^{4}\alpha- \left( \sin \left( \phi \right) 
 \right) ^{4}{r}^{2}- \left( \cos \left( \phi \right)  \right) ^{2}{b}
^{2} \right)  \left( \alpha\,{r}^{1-3\,\omega}-{r}^{2} \right) }}.
\end{eqnarray}
\end{widetext}
The Gaussian optical curvature $\mathcal{K}$, is defined as:
\begin{eqnarray}
\mathcal{K}=\frac{1}{\sqrt{det\bar{g}}}[\frac{\partial}{\partial\phi}
\left(\frac{\sqrt{det\bar{g}}}{\bar{g}_{rr}}\bar{\Gamma^{\phi}_{rr}}\right)-\frac{\partial}{\partial r}
\left(\frac{\sqrt{det\bar{g}}}{\bar{g}_{rr}}\bar{\Gamma^{\phi}_{r\phi}}\right)].
\end{eqnarray}
So by using Christopher symbols and the non-zero metric components, we get

\begin{eqnarray}
\mathcal{K}\approx-\frac{2M}{r^{3}}+\frac{3Q^{2}}{r^{4}}-\frac{6MQ^{2}}{r^{5}}+\frac{1}{2r^{6}}\left[-(9\omega^{2}+9\omega+2)r^{3-3\omega}+18M(\omega^{2}+\frac{2}{3}\omega+\frac{1}{3})r^{2-3\omega}9Q^{2}(\omega^{2}+
\frac{1}{3}\omega+\frac{2}{3})r^{1-3\omega}\right]\alpha \notag
\\+\frac{15aM H(r,\phi,\alpha)}{r^{9}}.
\end{eqnarray}
The last term gives the rotation. Where the function $H(r,\phi)$ is as follows
\begin{widetext}
\begin{eqnarray}
H(r,\phi,\alpha)&=&(r^{2}+a^{2}+Q^{2}-\alpha r^{1-3\omega})r^{-9}(r^{2}\sin^{4}\phi+
b^{2}\cos^{2}\phi)^{\frac{-7}{2}}\nonumber\\&\times&
[r^{10}\sin^{12}\phi-\frac{b^{2}r^{4}\sin^{10}\phi}{2}(r^{2}+a^{2}+Q^{2}-\alpha r^{1-3\omega})^{2}\nonumber\\&+&\frac{\cos^{2}\phi}{2}[-4r^{4}-71r^{2}(a^{2}+Q^{2}-
\alpha r^{1-3\omega})\frac{49}{11}\nonumber\\&-&10(a^{2}+Q^{2}-
\alpha r^{1-3\omega})^{2}]b^{2}r^{4}\sin^{8}\phi+4b^{3}\cos^{3}\phi\nonumber\\&\times&
(r^{2}+\frac{1}{2}(a^{2}+Q^{2}-\alpha r^{1-3\omega}))
(r^{2}+a^{2}+Q^{2}-\alpha r^{1-3\omega})\nonumber\\&\times&r^{3}\sin^{7}\phi+b^{2}\cos^{2}
\phi(2b^{2}-5r^{2}\cos^{2}\phi)(r^{2}+a^{2}\nonumber\\&+&
Q^{2}-\alpha r^{1-3\omega})^{2}r^{2}\sin^{6}\phi+8b^{3}\cos^{4}\phi(r^{2}+\frac{1}{2}(a^{2}+Q^{2}
\nonumber\\
&-&\alpha r^{1-3\omega}))(r^{2}+a^{2}+Q^{2}-\alpha r^{1-3\omega})r^{3}\sin^{5}\phi-3b^{6}\nonumber\\&\times&
\cos^{6}\phi(r^{2}-\frac{1}{3}(-a^{2}-Q^{2}-\alpha r^{1-3\omega}))+\cos^{4}\phi(\frac{-9}{2}r^{2}
\nonumber\\&+&\frac{8}{3}r^{2}-\frac{7}{2})b^{2}r^{2}
\sin^{4}\phi+5b^{4}\cos^{6}\phi(r^{2}+a^{2}+Q^{2}\nonumber\\
&-&\alpha r^{1-3\omega})^{2}r^{2}\sin^{2}\phi-2b^{5}r^{3}\cos^{6}\phi(r^{2}+4(a^{2}+Q^{2}\nonumber\\&-&\alpha
r^{1-3\omega}))\sin\phi-r^{3}b^{5}
\cos^{4}\phi\sin^{3}\phi((r^{2}+4(a^{2}+Q^{2}\nonumber\\&-&\alpha r^{1-3\omega}))].\nonumber
\end{eqnarray}
\end{widetext}
Where the Gaussian optical curvature depends on the BH parameters, $a, M, \\Q^{2},\alpha$.
In the next part, by using Gaussian optical curvature we compute angle of deflection.
\section{Deflection Angle Using Gauss Bonnet Theorem}
Let the domain $(D_{R}, \bar{g})$ be a simply connected over the osculating Riemannian
manifold $(\mathcal{M}, \bar{g})$ along the boundaries of $\tilde{\gamma}_{\bar{g}}$ and
$\tilde{\gamma}_{\tilde{R}}$. Furthermore, we define Gaussian curvature as a $\mathcal{K}$
of $(\mathcal{M},\bar{g})$ and geodesic curvature represents as $\kappa$
of $\partial D_{\tilde{R}}=\tilde{\gamma}_{\bar{g}} \cup \tilde{\gamma}_{\tilde{R}}$. Thus by the Gauss-Bonnet theorem
\begin{equation}
\int\int_{D_{R}} K d\tilde{\sigma}+ \oint_{\partial D_{R}} \kappa dx + \sum_{n} \Theta^{n}= 2 \pi \mathcal{X} (D_{R}).
\end{equation}

Here $d\tilde{\sigma}$ is defined as the surface element, $\Theta_{n}$ is the exterior angle at the $nth$ vertex,
$\mathcal{X}(D_{R})$ be defined as Euler characteristic number. The geodesic curvature use to determine the
deviation from geodesic. Since in case of geodesics $\tilde{\gamma}_{\bar{g}}$ the geodesic curvature vanishes
i-e $\kappa(\tilde{\gamma}_{\bar{g}})=0$, if an only if, $\tilde{\gamma}$ is geodesic. Now, our main focuss on
calculating $\kappa(\tilde{\gamma}_{\bar{g}})$. Which is defined as
\begin{equation}
\kappa(\tilde{\gamma}_{R})=\mid\nabla_{\dot{\tilde{\gamma}}R}\dot{\tilde{\gamma}}\mathcal{R}\mid.
\end{equation}
Now we consider $\tilde{\gamma}_{R} :=r(\phi)= R = const$, in such a way the radial part states that
\begin{equation}
(\nabla_{\dot{\tilde{\gamma}}R}\dot{\tilde{\gamma}}\mathcal{R})^{r}=
\dot{\tilde{\gamma}}^{\phi}_{R}(\partial_{\phi}\dot{\tilde{\gamma}}^{r}_{R})+
\bar{\Gamma}^{r}_{\phi\phi}(\dot{\tilde{\gamma}}^{\phi}_{R})^{2}.
\end{equation}
It is obvious that the first term vanishes, where by using the unit speed condition $(\bar{g}_{\phi\phi}\dot{\tilde{\gamma}}^{\phi}_{R}\dot{\tilde{\gamma}}^{\phi}_{R})=1$
we find the second term. Seeing that our optical geometry is asymptotically Euclidean
for geodesic curvature, At $R\rightarrow \infty$ the geodesic curvature reduces to
$\kappa(\tilde{\gamma}_{R})\rightarrow \frac{1}{R}$. If we consider $R\rightarrow \infty$
the jump angles $(\Theta_{O},\Theta_{D})$ become $\pi$, in other words we say
that the sum of jump angles to the source of light $D$ and the position of observer $O$ satisfies
$\Theta_{O} + \Theta_{D}\rightarrow \pi$. Now, for constant $R$ the optical metric yields
\begin{eqnarray}
\lim_{R\rightarrow \infty}dx&=&\lim_{R\rightarrow \infty}\left[\sqrt{\frac{r^{4}}{\Delta_{r}
(\Delta_{r}-a^{2})}\left(\frac{dr}{dt}\right)^{2}}
\right.\nonumber\\&\times&\sqrt{\frac{(r^{2}+a^{2})^{2}-a^{2}
\Delta_{r}}{\Delta_{r}-a^{2}}+\frac{a^{2}((r^{2}+a^{2})+\Delta_{r})^{2}}{\Delta_{r}-a^{2}}
\left(\frac{d\phi}{dt}\right)^{2}}\nonumber\\
&-&\left.a\left(\frac{(r^{2}+a^{2})+\Delta_{r}}{\Delta_{r}-a^{2}}\right)\right]d\phi
\nonumber\\&\rightarrow& Rd\phi,
\end{eqnarray}
On the other hand we use
\begin{equation}
\lim_{R\rightarrow \infty}h(R)\rightarrow 1.
\end{equation}
Hence, we represent that
\begin{equation}
\lim_{R\rightarrow \infty}\kappa (\tilde{\gamma}_{R})\frac{dx}{d\phi}\rightarrow 1.
\end{equation}

By supposition, the source and observer are in the asymptotically Euclidean region,
consequently the last equation plainly shows our conclusion that our  optical metric
is asymptotically Euclidean. So by Gauss-Bonnet theorem geodesic curvature yields that
\begin{equation}
\int\int_{D_{R}}\mathcal{K}dD+\oint_{\tilde{\gamma}_{R}}\kappa dx=^{R\rightarrow 
\infty}\int\int_{D_{R}}\mathcal{K}d\tilde{\sigma}+\int^{\pi+\hat{\alpha}} _{0}d\phi =\pi,
\end{equation}
hence, we obtain
\begin{equation}
\Theta=-\int\int_{D_{\infty}} \mathcal{K}d\tilde{\sigma}.
\end{equation}
Now by using the straight line approximation of the light ray $(r=\frac{b}{\sin\phi})$,
we get the total deflection angle

\begin{eqnarray}
\Theta&\simeq& \frac{4M}{b}+\frac{2\alpha}{b}-\frac{3 \alpha \omega}{b}-\frac{3 Q^2}{4 b^2}\pm\frac{4Ma}{b^{2}}+\mathcal{O}( M),
\end{eqnarray}
where the positive and negative signs denoted the reversed and rotational movement of light rays. Our proposed deflection angle increases by increasing the mass term and quintessence parameter but as we increase the impact parameter, deflection angle decreases.
\section{Null Geodesics}
Now, by using the variational principle $\delta\int\mathcal{L}dD=0$ \cite{R13}, we find the deflection angle.
For this we set $\theta=\frac{\pi}{2}$ and $\dot{\theta}=0$, then we find the following equation
for lagrangian
\begin{eqnarray}
2\mathcal{L}&=&-(\frac{\Delta_{r}-a^{2}}{\Sigma^{2}\Xi^{2}})\dot{t}^{2}+\left(\frac{(r^{2}+
a^{2})^{2}-a^{2}\Delta_{r}}{\Sigma^{2}\Xi^{2}}\right)\dot{\phi}^{2}
+\frac{\Sigma^{2}}{\Delta_{r}}\dot{r}^{2}\nonumber\\&-&\left(\frac{2a((r^{2}+a^{2})-\Delta_{r})}
{\Sigma^{2}\Xi^{2}}\right)\dot{\phi}\dot{t},
\end{eqnarray}
furthermore $\dot{p}=\frac{d p}{d\tau}$ and $\tau$ stands for affine parameter along the geodesic.
Since the metric coefficients do not depend on cyclic coordinates $(t,\phi)$, then the conjugate
momenta of these cyclic coordinates are represented by $(\Omega_{t}, \Omega_{\phi})$ and are
preserved. Then the equations of motion are derived from $\dot{\Omega}_{p}-\frac{\partial\mathcal{L}}{\partial p}=0$ leading to
\begin{equation}
\dot{\Omega_{t}}=0,  \dot{\Omega_{\phi}}=0,
\end{equation}
here $\Omega_{p}=\frac{\partial\mathcal{L}}{\partial\dot{p}}$ are conjugate momenta to the
space-time coordinate $p$, and for lagrangian represented by
\begin{eqnarray}
\Omega_{t}&=&-\frac{\Delta_{r}}{r^{2}\Xi^{2}}\dot{t}-a\frac{(r^{2}+a^{2}-\Delta_{r})}{r^{2}\Xi^{2}}\dot{\phi}\equiv-F,\\
\Omega_{r}&=&\frac{r^{2}}{\Delta_{r}}\dot{r}^{2},\\
\Omega_{\phi}&=&-a\frac{(r^{2}+a^{2}-\Delta_{r})}{r^{2}\Xi^{2}}\dot{t}+\frac{r^{2}+a^{2}}{\Xi^{2}}\dot{\phi}\equiv N,
\end{eqnarray}
where $F$ and $N$ are constants of motion for a test particle, relating to the energy at infinity
and the angular momentum, respectively. Thus the Hamiltonian for a test particle is given by \cite{j1}
\begin{equation}
\mathcal{H}=\Omega_{t}\dot{t}+\Omega_{\phi}\dot{\phi}+\Omega_{r}\dot{r}-\mathcal{L},
\end{equation}
\begin{equation}
2\mathcal{H}=-F\dot{t}+N\dot{\phi}+\frac{r^{2}}{\Delta_{r}}\dot{r}^{2}\equiv-M^{2}.
\end{equation}
Now, we consider only massless particles and taking the condition\\$M^{2}=0$ for photons.
Then the geodesic equations of motion in the following way \cite{j2}
\begin{eqnarray}
\dot{t}&=&(-Fr^{2}+a N(1-\frac{1}{r^{2}}(a^{2}+Q^{2})+\alpha r^{-1-3\omega}))
(r^{2}+a^{2}\nonumber\\&+&Q^{2}-\alpha r^{1-3\omega})^{-1},\\
\dot{\phi}&=&(N+aF(a^{2}+Q^{2}-\alpha r^{1-3\omega}))(r^{2}+a^{2}+Q^{2}-\alpha r^{1-3\omega})^{-1},\\
\dot{r}^{2}&=&(-r^{2}F^{2}+a F N\left(1-\frac{1}{r^{2}}(a^{2}+Q^{2})
+\alpha r^{-1-3\omega}\right)\nonumber\\&-&r^{2}N(N+aF))
\left(1+\frac{1}{r^{2}}(a^{2}+Q^{2})-\alpha r^{-1-3\omega}\right)^{-2}.
\end{eqnarray}
The closest distance $r_{0}$ for the metric (\ref{J1}) can be calculated by taking $\dot{r}=0$, such that
\begin{equation}
\frac{r_{0}}{b}= \sqrt{1-\left(\frac{a}{b}\right)^{2}+r_{0}^{2}\left(1-\frac{a}{b}\right)^{2}(R^{2}-2M)},
\end{equation}
where
\begin{equation}
R^{2}=a^{2}+Q^{2}-\alpha r^{1-3\omega}.\nonumber
\end{equation}
Then, for bending angle
\begin{equation}
\Theta= 2 \int_{r_{0}}^{\infty} \mid\frac{d\phi}{dr}\mid dr -\pi,
\end{equation}
after simplification we get
\begin{eqnarray}
\Theta&\simeq& \frac{4M}{b}+\frac{2\alpha}{b}-\frac{3 \alpha \omega}{b}-\frac{3 Q^2}{4 b^2}\pm\frac{4Ma}{b^{2}}+\mathcal{O}( M),
\end{eqnarray}
now we utilize the change of variables such as $w= r_{0}/r$; Hence we suppose $r_{0}\simeq b$.

\section{Graphical Analysis}
This section is devoted to discuss the graphical behavior of deflection angle $\Theta$.
We also demonstrate the impact of BH charge $Q$,
quintessence $\alpha$, rotation $a$ and impact $b$ parameters
on deflection angle by examining the stability and instability of BH.
\subsection{Deflection angle with Impact parameter $b$}
This subsection is based on the analysis of deflection angle $\Theta$ with impact parameter $b$
for different values of rotation parameter $a$, BH charge $Q$,
and quintessence parameter $\alpha$ for fixed $M=1$ and $\omega=-\frac{2}{3}$.
\begin{center}
\epsfig{file=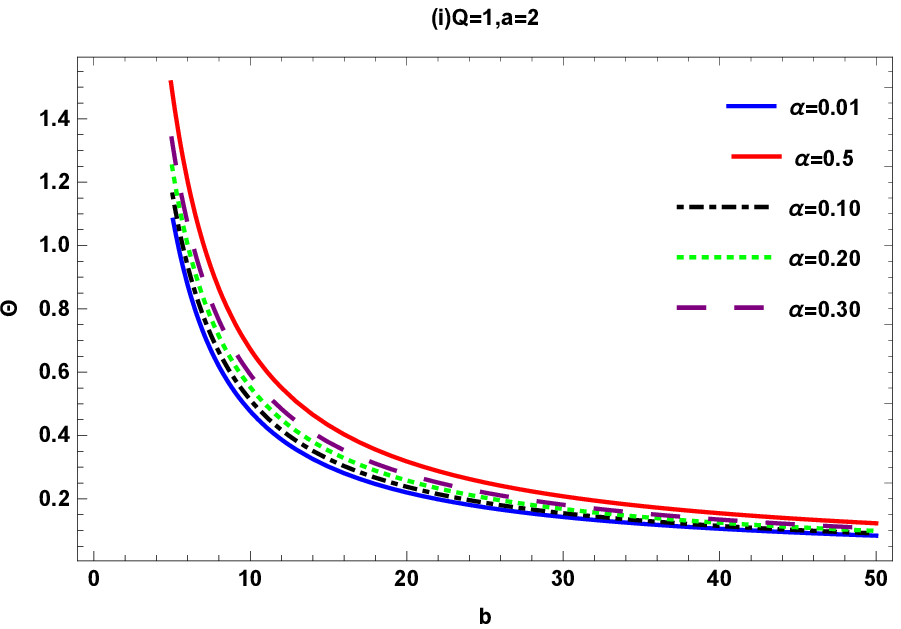,width=0.50\linewidth}\epsfig{file=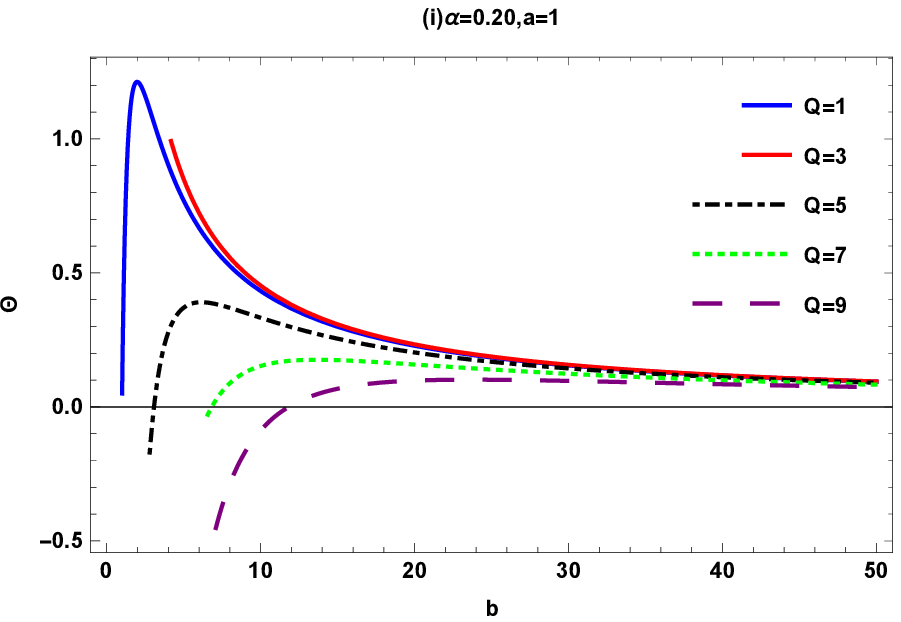,width=0.50\linewidth}\\
{Figure 2: Relation between $\Theta$ and $b$}.
\end{center}
\begin{itemize}
\item \textbf{Figure 2} shows the behavior of $\Theta$ w.r.t impact parameter $b$.
\begin{enumerate}
\item In figure (i), This plot shows the behavior of $\Theta$ with $b$ for fixed $Q$,
 $a$ and varying $\alpha$. We analyzed that deflection angle gradually 
 decreasing and then goes to positive infinity. We analyzed the stable behavior of
 deflection angle only for positive values of quintessence parameter.
 \item In figure (ii), This plot represents the behavior of $\Theta$ with $b$ for fixed $\alpha$,
  $a$ and varying $Q$ we observed that the deflection angle initially increasing from negative to positive
   and then gradually decreasing which shows the stable behavior.
\end{enumerate}
\end{itemize}
\begin{center}
\epsfig{file=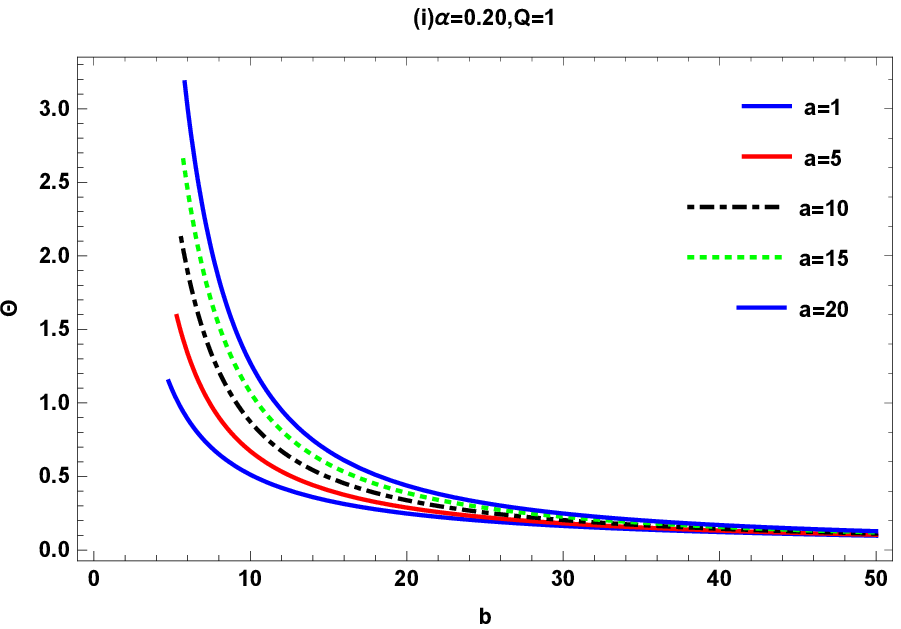,width=0.50\linewidth}\\
{Figure 3: Relation between $\Theta$ and $b$}.
\end{center}
\begin{itemize}
\item \textbf{Figure 3} shows the behavior of $\Theta$ w.r.t impact parameter $b$.\\
  This figure shows the behavior of $\Theta$ with $b$ for fixed $Q$, $\alpha$ and varying
   $a$. We observed that for negative value of $a$ the deflection angle is negatively decreasing
   but for $a>0$ deflection angle initially exponentially
   decreasing and then goes to positive infinity.
\end{itemize}
\section{Conclusion}
In this paper, we have analyzed the weak gravitational lensing
in the background of Kerr-Newman BH with quintessential dark energy.
Initially, we have found the deflection angle of light by slowly
rotating charged BH with quintessence by means of GBT. For this
purpose, firstly we have consider the photon rays into equatorial
plane. Afterthat we have obtained the corresponding optical metric
and calculated the Gaussian optical curvature in leading order
terms. Consequently, we have calculated the deflection angle of light
by slowly rotating charged BH with quintessence. We also investigated 
the graphical behavior of impact parameter on deflection angle.\\
\textit{\textbf{Deflection angle by slowly rotating Charged BH with Quintessence with respect to Impact parameter:}}
\begin{itemize}
\item The behavior of $\Theta$ w.r.t $b$ can be analyzed.

\item For particular value of $b$, we can only obtain the stable behavior for $0.01\leq\alpha \leq1$.

\item We observe that for the variation of $Q$ the deflection angle gradually decreasing,
 which indicates that the BH evolution appears initially from physical stable state to unstable.

\end{itemize}
In addition, we have also obtained the deflection angle of light by Kerr-Newman BH. Moreover,
for verification of our results we have used the null geodesics
method and calculated the deflection angle of light by Kerr-Newman
BH with quintessence. It is noted that the correspondence between the deflection angle obtained
by GBT and geodesic method is exact up to the first order terms,
while this correspondence fails going to higher order terms \cite{j2}
the reason behind this is the straight line approximation engaged in
the integration domain. In principle, this incompatibility can be
overcome by choosing a more precise relation for the light ray
trajectory in the integration domain. It is worth mentioning that,  our proposed deflection angles are reduced to the Einstein
deflection angle $\Theta=4M/b$ upto the first order terms
and which is not influenced by the electric charge and if we take $\alpha=Q=0$
then our results reduced into Kerr solution.
The Kerr-Newman BH solution may be applicable in astrophysics.
The results we analyzed from the analysis of deflection angle $\Theta$ by Kerr-Newman
BH with quintessence are generalized as follows:\\
\textit{\textbf{Deflection angle by Kerr-Newman BH with quintessence with respect to Impact parameter:}}
\begin{itemize}
\item The behavior of $\Theta$ w.r.t $b$ can be analyzed for the particular range of $b$.

\item We can analyzed the deflection angle by Kerr-Newman BH with quintessence for $0.01\leq\alpha \leq1$.

\item We observe that for the variation of $Q$ the deflection angle goes from negative
 to positive and then decreasing which shows that the deflection angle initially at non-physical 
 unstable state and then goes to stable stable.

\item For positive values of $a$, the deflection angle exponentially decreases then become
 identical. Moreover, we can say that $a > 0$ indicates that BH is co-rotating regarding to the observer, whereas for $a < 0$ it is relating to counter-rotating BH.
\end{itemize}

\end{document}